\documentclass[prl,twocolumn,superscriptaddress]{revtex4}

\usepackage{hyperref}
\usepackage{amsmath,amssymb,graphicx}
\usepackage{color}

\newcommand{\sslash}{\mathbin{/\mkern-4mu/}}

\graphicspath{{fig-ps/}}

\begin{document}

\date{\today}

\title{Robust Vortex Lines, Vortex Rings and Hopfions in 3D Bose-Einstein Condensates}

\author{R.N.~Bisset}
\email{rnbisset@gmail.com}
\affiliation{Center for Nonlinear Studies and Theoretical Division, Los Alamos
National Laboratory, Los Alamos, NM 87545}

\author{Wenlong Wang}
\affiliation{Department of Physics, University of Massachusetts,
Amherst, Massachusetts 01003 USA}

\author{C.~Ticknor}
\affiliation{Theoretical Division, Los Alamos
National Laboratory, Los Alamos, NM 87545}

\author{R.~Carretero-Gonz{\'a}lez}
\affiliation{Nonlinear Dynamical Systems
Group,
Computational Sciences Research Center, and
Department of Mathematics and Statistics,
San Diego State University, San Diego, California 92182-7720, USA}

\author{D.J.~Frantzeskakis}
\affiliation{Department of Physics, University of Athens,
Panepistimiopolis, Zografos, Athens 15784, Greece}

\author{L.A.~Collins}
\affiliation{Theoretical Division, Los Alamos
National Laboratory, Los Alamos, NM 87545}

\author{P.G.~Kevrekidis}
\affiliation{Department of Mathematics and Statistics, University of Massachusetts,
Amherst, Massachusetts 01003-4515 USA}

\affiliation{Center for Nonlinear Studies and Theoretical Division, Los Alamos
National Laboratory, Los Alamos, NM 87545}

\begin{abstract}

Performing a systematic Bogoliubov-de Gennes spectral analysis, 
we illustrate that stationary vortex lines, vortex rings and more exotic states, such as hopfions, 
are robust in three-dimensional atomic Bose-Einstein condensates, for large parameter intervals.
Importantly, we find that the hopfion can be stabilized in a simple
parabolic trap, without the need for trap rotation or inhomogeneous interactions.
We supplement our spectral analysis by studying the dynamics of such stationary states; 
we find them to be robust against significant perturbations of the initial state.
In the unstable regimes, we not only identify the unstable mode, such as a quadrupolar or hexapolar mode, 
but we also observe the corresponding instability dynamics.
Furthermore, deep in the Thomas-Fermi regime, we investigate the particle-like behavior of 
vortex rings and hopfions.

\end{abstract}

\maketitle

\maketitle


{\it Introduction:} 
Atomic Bose-Einstein condensates (BECs) have offered, over the last two decades,
a fertile playground for the exploration of nonlinear matter 
waves~\cite{book1,book2,emergent}.
While a large volume of the early
work along this vein focused on 
solitons and vortices, the 
remarkable advancement of computational resources
has rendered more accessible the frontier of three-dimensional (3D) structures.
Arguably, the most prototypical among the latter, not only in 
superfluid but also in regular fluid settings~\cite{saffman,Pismen1999},
is the vortex ring (VR). VRs have not only been theoretically predicted,
but also experimentally observed 
(see the reviews \cite{emergent,komineas_rev,book_new}).

In addition to the VRs and vortex lines (VLs) extensively studied
in earlier BEC experiments 
(see, e.g., 
Refs.~\cite{fetter1,fetter2} and a 
more recent experimental realization in Ref.~\cite{dshall}), BECs may support 
more complex topological structures, such as Skyrmions in multi-component settings.
In its simplest realization, originally proposed in Refs.~\cite{Ruostekoski01,ruost}, 
the Skyrmion consisted of a VR in one-component,
coupled to a VL in the second component.
Interestingly, more 
complex Skyrmion states involving three-component spinor
BECs were realized experimentally in both two-~\cite{shin} and 
three-dimensions~\cite{bigelow},
involving respectively, coupled states of topological charge
$S=-1, 0, 1$ and $S=0, 1, 2$, and described in the recent
theoretical work of Ref.~\cite{ruost2}. 
Of increasing interest of late is the one-component 
counterpart to the Skyrmion, namely the so-called hopfion state~\cite{boris,yakimenko}.
This state consists of a VR and VL in the same component, with the 
axis of the VR coinciding with that of the VL.
%
A stable hopfion state has so far only been predicted in somewhat 
complicated experimental configurations.  These include, for instance, 
elaborated radially increasing nonlinear interactions~\cite{boris},  
or a rotation of the trap~\cite{yakimenko}.
Here we show that the hopfion can, in fact, be stable 
for large chemical potential ranges and simple trapping configurations,
i.e., inside a parabolic trap.

Our main aim is to provide a systematic stability analysis of
VRs, hopfions and, in passing, VLs.
By a detailed understanding of the pertinent
modes of the Bogoliubov-de Gennes (BdG) linearization,
we are able to explain when the relevant stationary states 
are stable or unstable.
We also elucidate, both, how such properties depend on geometric characteristics
such as the trap aspect ratio, and what instabilities one may encounter
in different parameter intervals.
These results should pave the way for the experimental identification 
of such coherent structures in state-of-the-art experimental setups of 3D BECs.
%

In what follows, we first discuss the stability properties of the VR from an analytical perspective.
We then corroborate these analytical predictions of the relevant modes
by means of highly-intensive numerical spectrum computations.
We also show
how the stability of the VR and the VL implies the potential stability of 
the hopfion pattern, and confirm this with our numerics.
%
Finally, we provide an exploration of the unstable dynamics of the VR and
hopfion states informed by both spectral properties and direct numerical simulations.


{\it Analytical Considerations:} 
We begin with the 3D Gross-Pitaevskii equation (GPE):
\begin{equation}
i \psi_t=-\frac{\hbar^2}{2m} \nabla^2 \psi +V(r) \psi +g| \psi |^2 \psi,
\label{GPE}
\end{equation}
where $\psi$ is 
the wavefunction of the 3D Bose-gas near zero temperature, 
$g=4\pi\hbar^2a_s/m$, with $a_s$ being the $s$-wave scattering length and $m$ is the particle mass.
The potential assumes the prototypical form of the harmonic oscillator,
$V(x,y,z)= m\omega_r^2 r^2/2 + m\omega_z^2 z^2/2$,
with $\omega_r$ and $\omega_z$ being the
planar and transverse trapping strengths, respectively.
The case of $\omega_z>\omega_r$ leads to an oblate BEC, while the reverse
inequality produces a prolate BEC.

In earlier work we explored the bifurcation of a
VR near the linear limit of low density, either from a planar or
from a ring dark soliton~\cite{pra_all}; here, our focus will be on the
opposite limit. In particular, we consider the case of large
chemical potential (and small healing length) where the 
VRs can be considered as particle-like objects in their behavior
and dynamics. Reference~\cite{fetterpra}, following
the pioneering work of Ref.~\cite{Roberts71}, 
explored the dynamics of the single VR
 in the presence of a trap. This is also examined in Ref.~\cite{horng},
where the results of Ref.~\cite{fetterpra} are utilized. In particular,
in Ref.~\cite{fetterpra}, the expression for the velocity of a vortex
line element is given by
\begin{eqnarray}
{\bf v}({\bf x}) = \Lambda \left(\kappa \hat{{\bf b}} +
\frac{\hat {\bf t} \times \nabla V}{F(r,z)} \right) ,
\label{vel}
\end{eqnarray}
where 
$\Lambda=(-1/2) {\rm ln}({\sqrt{R_r^{-2}+ \kappa^2/8}}/\sqrt{2 \mu})$
and
$\kappa$ denotes the curvature of the element (for a VR
$\kappa=1/r$). We denote by
$\hat{{\bf b}}$ the binormal vector
(for the axisymmetric VR, it is $\hat{{\bf z}}$), while
$\hat {\bf t}$ denotes the tangent vector (equal to
$\hat {\bf \theta}$ for the VR). The quantity
$F(r,z)=\max\{\mu - V(r),0\}$, represents the Thomas-Fermi (TF) 
density profile, relevant to the case of large density, 
equivalent to large chemical potential, analyzed here; the corresponding 
 radial and axial 
 TF radii 
 are given by
$R_{r,z}=(2 \mu/\omega_{r,z}^2)^{1/2}$.
%
%
Note that $\Lambda$ is accurate up
to logarithmic corrections, 
which will be responsible
for the approximate nature of the analytical results (in comparison
to the numerical findings) in what follows.

Assuming that $\Lambda(r)$ varies slowly with $r$ (indeed logarithmically),
the following equations of motion are then derived for a single VR
inside the trap~\cite{horng}:
\begin{eqnarray}
\frac{1}{\Lambda(r)} \dot{r} =\frac{ \omega_z^2 z}{F(r,z)},
\quad \dot{\theta}=0, \quad \frac{1}{\Lambda(r)} \dot{z}=
\frac{-\omega_r^2 r}{F(r,z)} + \frac{1}{r} .
\label{singlering}
\end{eqnarray}
These equations predict the presence of an
equilibrium radius of the VR in the $z=0$ plane, i.e.~$r_{\rm eq}=\sqrt{2 \mu/(3 \omega_r^2)}$.
This effective radius seems to provide a natural generalization
of the radius of the 
ring dark soliton~\cite{rds2003,kamch},
and is also in line with 
earlier results
~\cite{jackson}.

Reference \cite{horng} chiefly focused on considering the dynamics of 
azimuthal perturbations (Kelvin-wave type modulations) using Eq.~(\ref{vel}). In particular,
they linearized around
the stationary solution $(r,z)=(r_{\rm eq},0)$, using a perturbation
of the form $r(t)=r_{\rm eq} + R_1(t) e^{i n \phi}$ and
$z(t)=Z_1 e^{i n \phi}$, for integer $n$.
Substitution of this linearization ansatz in
Eq.~(\ref{vel}) provides a generalized set of equations [Eqs.~(5) and (6) in Ref.~\cite{horng}]
from which the effect of the azimuthal modulations on the motion of
the VR can also be evaluated.
Importantly, notice that this set of 
equations can account for the $n=0$ oscillatory motion of the VR inside
the trap, described by Eq.~(\ref{singlering}).
The final result that we 
explore numerically in what follows is that the frequencies of
vibration of the VR are given by,
\begin{eqnarray}
\omega=\pm \frac{3 \Lambda(r_{\rm eq}) \omega_r^2}{2 \mu} 
\left[ (n^2- \frac{\omega_z^2}{\omega_r^2}) (n^2-3) \right]^{1/2} .
\label{freq}
\end{eqnarray}
Importantly, we intend to test the ensuing implication that the VR stability depends on the shape of the condensate.
More specifically, if the condensate is 
prolate ($\omega_z/\omega_r < 1$), then the VR should be unstable due to 
the $n=1$ mode. If the condensate is spherical to slightly oblate
($1 \leq \omega_z/\omega_r  \leq 2$),
then at the particle level and under azimuthal perturbations,
the VR and VL should be stable. Finally, for sufficiently oblate condensates,
with $\omega_z/\omega_r>2$, the VR should be unstable due to the $n=2,
\dots, [\omega_z/\omega_r]$ modes, where the brackets denote the integer
part function.


{\it Numerical Results:} We start
by considering the spectral linearization analysis around a VR state. The relevant BdG ansatz 
will be of the form:
\begin{eqnarray}
\psi=e^{-i \mu t} \left[ \psi_0 + \epsilon \left(u(x,y,z) e^{i \omega t} 
+ v^{*}(x,y,z) e^{-i \omega^{*} t} \right) \right],
\label{bdg}
\end{eqnarray}
where $\mu$ is the chemical potential,
the asterisks denotes complex conjugation, $\epsilon$ is a formal small parameter,   
and the presence of imaginary (or complex) eigenfrequencies
$\omega$ reflects an instability along the direction of the corresponding eigenvector $(u,v)^T$.

In practice, we solve the GPE, Eq.~(\ref{GPE}), using a Newton-Krylov scheme \cite{Kelly2003}. 
For the BdG  equations we utilize the azimuthal symmetry of the trap in a way similar to 
Refs.~\cite{Ronen2006,Blakie2012}. This amounts to using spectral-basis modes that each have a definite 
angular momentum quantum number, $m$, proportional to $e^{im\phi}$. 
This way, we can treat the azimuthal variable analytically, effectively reducing the problem to 2D. 
Importantly, for a given excitation, the coupling between $m$-subspaces is limited, and this allows the 
diagonalization of relatively small subsets independently. Specifically, if the $v$ of Eq.~(\ref{bdg}) resides 
in subspace $m$, then the $u$ resides in subspace $m+2s$, where $s$ is the charge (angular momentum quantum 
number) of the stationary state \cite{Dodd1997}.

\begin{figure}[tb]
\begin{center}
\includegraphics[width = 3.4in]{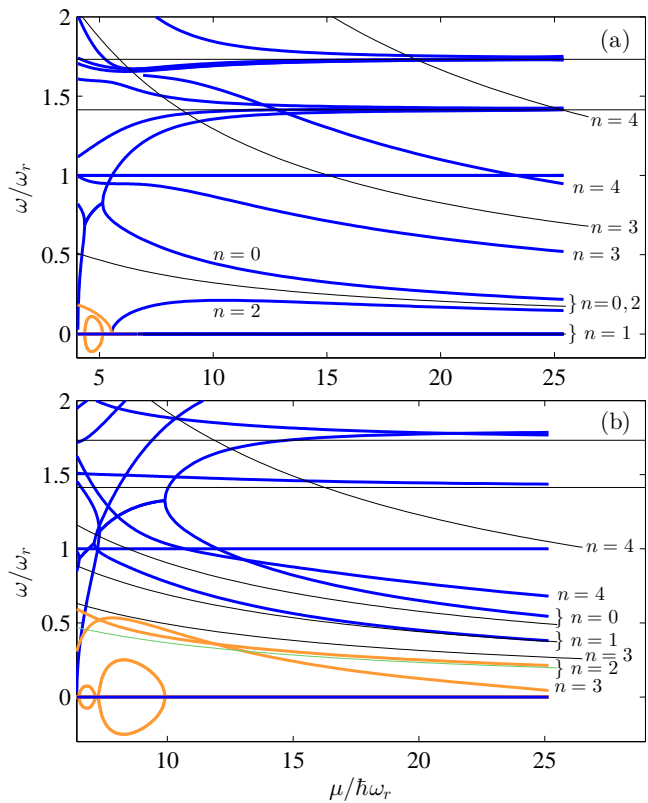}
\caption{
(Color online) BdG-frequency spectrum for the vortex ring 
when (a) $\omega_z/\omega_r=1$ and (b) $\omega_z/\omega_r=2.8$. 
Stable (real) components are depicted by the thick blue (dark) lines and 
unstable (imaginary) components by the thick orange (light) lines. 
Thin lines are TF analytic predictions [black (dark) corresponding to
real and green (light) to imaginary components of the eigenvalues] for the vortex ring 
[thin curved lines, Eq.~(\ref{freq})] and the ground state 
(thin horizontal lines; see Ref.~\cite{Stringari1996}). 
The undulation number $n$ [see Eq.~(\ref{freq})] of each excitation is indicated. 
For both geometries, the analytical prediction is in good agreement with the 
numerics for $n=0,1$ and $2$, but deteriorates for $n>2$.
}
\label{Fig:VRfz1}
\end{center}
\end{figure}

Our fundamental premise in what follows
is that for the large $\mu$ regime, the spectrum
of a state such as the VR or the VL consists of the union of two principal
ingredients: 
the modes of vibration (undulation) of the VR or VL itself, 
and the modes of the underlying ``ground state'' 
(see Fig.~\ref{Fig:GSfz1-GSfz2p8} in the Supplemental Material), i.e., the TF cloud.
The TF spectrum in this limit has been identified early on \cite{Stringari1996} and the 
internal modes of the VR are described by Eq.~(\ref{freq}).
Hence, in the numerical
BdG spectra of Fig.~\ref{Fig:VRfz1}(a) for $\omega_z/\omega_r=1$
and Fig.~\ref{Fig:VRfz1}(b) for $\omega_z/\omega_r=2.8$, we expect to observe
this union of modes. We find that this expectation is indeed
confirmed in both of these figures. The TF vibrational modes predicted 
by Ref.~\cite{Stringari1996} correspond
to the constant frequencies (thin horizontal lines) 
to which the relevant spectral
modes approach asymptotically. On the other hand, the modes predicted by
Eq.~(\ref{freq}) feature a decay as $1/\mu$ (modulated by 
a logarithmic dependence).
For the internal 
VR modes, the theoretical predictions, given by thin curved lines,
reasonably approximate the numerical results, especially for larger values of $\mu$. 
This approximation is especially good
for the anomalous-vibration mode of the entire VR as a whole inside
the trap ($n=0$), as well as for undulations of the VR such as the
dipolar ($n=1$) and quadrupolar ($n=2$) modes. 
We note, however, that for higher-order modes
(larger values of $n$), the analytical prediction of their
frequency, based on Eq.~(\ref{freq}), is less accurate.
The case of $\omega_z/\omega_r=1$ is expected to be stable in the large $\mu$ limit, 
as predicted by Eq.~(\ref{freq}), and
indeed this is supported by our BdG analysis in Fig.~\ref{Fig:VRfz1}(a).
On the other hand, for 
$\omega_z/\omega_r=2.8$, our results confirm that the VR
is unstable due to the mode of $n=2$, as was predicted by Eq.~(\ref{freq}) 
--- see Fig.~\ref{Fig:VRfz1}(b).
However, the mode of $n=3$ is
predicted to be stable analytically, while it is found to be unstable
numerically, illustrating the limitations of the analytical prediction for $n>2$.

Similar results can be found for the VL case 
(see Fig.~\ref{Fig:GSVL} in the Supplemental Material).
In this case, the analytical predictions could be suitably extended
by taking the curvature to be $0$, and accounting for $F(r,z)$ 
in Eq.~(\ref{vel}) not being a constant. A nontrivial difference between the latter
and the former case is that the VL is found to be generically stable
for the isotropic limit of $\omega_z/\omega_r=1$, i.e., the intervals
of instability due to imaginary or complex eigenvalues observed
in Fig.~\ref{Fig:VRfz1} are absent. The VR on the other hand features,
for small $\mu$, the unstable modes described in detail in Ref.~\cite{pra_all}.

\begin{figure}[tb]
\begin{center}
\includegraphics[width = 3.5in]{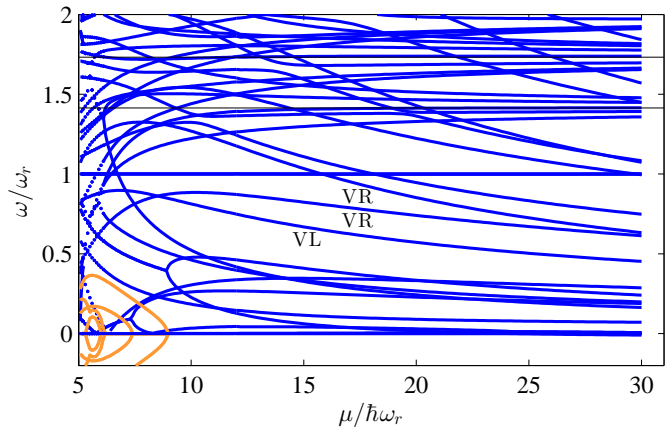}
\caption{
(Color online) The spectrum of the hopfion with $\omega_z/\omega_r=1$. 
The different types of curves have the same meaning as in Fig.~\ref{Fig:VRfz1}.
Importantly, we predict that the hopfion is stable for $\mu> 9$.
While some excitations can be identified as strictly belonging to either the VR or the VL, see labeled examples, other excitations correspond to hybrid modes due to the coupling between the VR and the VL.
}
\label{S3}
\end{center}
\end{figure}

Having identified stable VR and VLs in the isotropic limit for
large $\mu$ suggests that the hopfion itself, consisting of a combined VR and VL in the
same BEC, is likely to be stable in the TF limit. 
We have tested this prediction for $\omega_z/\omega_r=1$ in Fig.~\ref{S3}: 
indeed, we observe 
that although instabilities may arise for small values of $\mu$, for large
values of $\mu$ the hopfion is robust.
By investigation of the individual BdG eigenvectors we find that the spectrum 
encompasses the union of VR, VL and TF modes, with the analytical prediction for the latter 
\cite{Stringari1996} shown as 
thin horizontal lines. We have also labeled a few examples of purely VR and VL modes. 
However, we note that other modes demonstrate coupling between the VR and VL, 
as evidenced also by the hybrid nature of their BdG eigenvectors.


\begin{figure}
\begin{center}
\includegraphics[width = 3.5in]{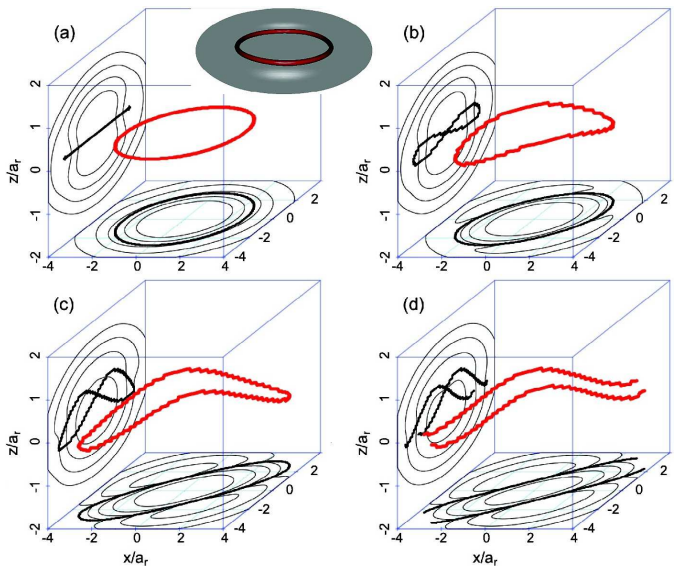}
\caption{
(Color online) Dynamics of the VR, demonstrating instability to an $n=2$ undulation, 
shown at times $\omega_rt = $ 3.5 (a), 25.4 (b), 29.0 (c) and 30.0 (d). 
Parameters: $\omega_z/\omega_r=2.8$ and $\mu/\hbar\omega_r \approx 15.3$.
 Red (gray) curves indicate vortex-core positions; thick black lines line show the vortex-core projections onto the $(x,y)$ and $(y,z)$ planes; thin-black contour lines represent the column density projections at 0.2, 0.4, 0.6 and 0.8 of the peak. Inset: iso-density surface of the stationary state at 0.1 of the peak; the inner iso-density surface of the VR is colored red. Units are that of the harmonic oscillator $a_r = \sqrt{\hbar/m\omega_r}$.
%
To see a movie of the dynamics see Ref.~\cite{movies} [Movie\#1].
}
\label{Fig:VRm2Dyn}
\end{center}
\end{figure}

To complement our spectral analysis, we explore the nonlinear dynamics
of the VR and the hopfion, utilizing 
two methods.
In the first, we temporally propagate the time-dependent GPE by employing a 
real-space product scheme, based on a split-operator approach, with the spatial 
component treated with a finite-element discrete-variable representation, 
using a Gauss-Legendre quadrature within each element \cite{Schneider2006}.
For the second method, we use a split-step operator on an FFT grid. 
For the results presented herein, we find quantitative agreement 
between these two numerical methods.

We first discuss the stability of the hopfion. At $\mu=12$, where according to Fig.~\ref{S3} the hopfion is stable, we performed two stability tests. In the first, we added an average of 1\% random noise to $\psi_0$ on each grid point at time $\omega_rt=0$, and found that the hopfion remains robust for the entire duration of the test, up to time $\omega_rt=200$. In the second test, we excited the hopfion by adding a special excitation $\psi_1$ at the 5\% level, i.e.~$\psi_0\to\psi_0+0.05\psi_1$, and found the hopfion to undulate in a stable manner for more than $\omega_rt=200$ time units. 
Note that $\psi_1$ is the 
most unstable mode for small chemical potentials, $\mu/\hbar\omega_r<9$ (see Fig.~\ref{S3}).


We now 
consider the instability dynamics. 
In Fig.~\ref{Fig:VRm2Dyn}, 
we examine the dynamics of the VR in a regime where it is found to
be dynamically unstable in our earlier spectral analysis, in particular
due to the $n=2$ quadrupolar mode. As a result, we observe 
that the relevant mode (i.e., the 
Kelvin wave) is amplified, until it eventually gives rise to the ``rupture'' of
the VR into a pair of VLs. We have also observed similar
dynamical evolutions where the instability is seeded by the unstable $n=3$ mode 
(see Fig.~\ref{Fig:VRN3Dyn} in the Supplemental Material).
In this case, we found that the VR breaks into 6 VLs, before 
exhibiting a temporary revival of a smaller VR.

\begin{figure}[tb]
\begin{center}
\includegraphics[width = 3.5in]{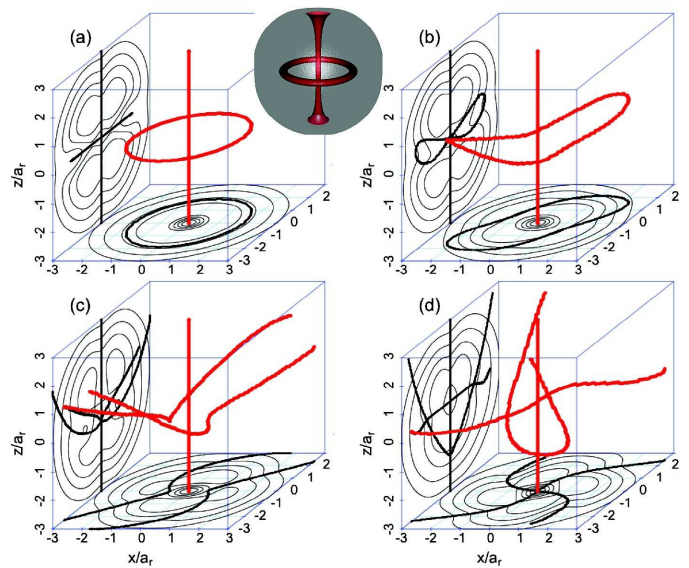}
\caption{
(Color online) Dynamics of a hopfion instability at times 
$\omega_rt = $ 0 (a), 40.9 (b), 42.4 (c) and 44.3 (d), 
after a random-noise seeding at $\omega_rt =0$. Parameters: $\omega_z/\omega_r=1$ and $\mu/\hbar\omega_r \approx 7.0$. Curves and contours have the same meanings as in Fig.~\ref{Fig:VRm2Dyn}.
To see a movie of the dynamics see Ref.~\cite{movies} [Movie\#3].
}
\label{Fig:HopfDyn}
\end{center}
\end{figure}

Finally, we examine the dynamical instability of the hopfion for 
$\mu\approx7$, where it is still dynamically unstable prior to
its stabilization for larger $\mu$.
The initial stages of the hopfion instability proceed in a similar manner to that of the VR. 
First, the VR part of the hopfion bends and then breaks into two
VLs. The main difference arises when these 
VLs are subsequently pulled inwards to reconnect with the vertical VL.
Interestingly, the VLs remain connected, with a `+' junction, for extended periods 
[see Fig.\ref{Fig:HopfDyn}(d)].
Note that such reconnection events, particularly interesting
in their own right and especially relevant in turbulent dynamics
(see, e.g., Refs.~\cite{white,tsatsos}), have also been observed 
in the presence of rotation~\cite{yakimenko}.

It is important to point out that the regimes considered in this work are experimentally accessible:  
for 
$\mu/\hbar\omega_r=10.1$, a regime in which Fig.~\ref{S3} predicts the hopfion 
to be stable, corresponds to a $^{87}$Rb BEC containing $\approx 1.1\times 10^4$ atoms
in an isotropic trap with $\omega_z=\omega_r=2\pi \times50$~Hz. 

{\it Conclusions and Outlook.} In the present work, we investigated
3D states that are supported in isotropic (and anisotropic) BECs. 
We used highly intensive numerical computations to explore their 
spectral stability, and found that,  
in the isotropic limit, the vortex line, vortex ring and their combined state, 
the hopfion, are dynamically
stable in a wide parameter regime.
Importantly, we predict the hopfion to be robust in ``ordinary'' condensates, 
within typical parameter regimes of a highly accessible parabolic trap.
%
Not only did we identify this
stability, but we provided a road map on how to ``read'' the spectrum 
of these different states and what principal ingredients are contained 
therein. 
Both the analytical approach and the 
numerical computations explained why different trap aspect ratios 
may drastically affect the stability properties of such states.
Finally, when the states were
deemed to be unstable, direct numerical simulations elucidated their
breakup and subsequent dynamics, such as the vortex line
reconnections in the case of the hopfion.


An interesting future direction would be the extension 
of our investigations to a higher number of components: 
in particular, in the two-component setting, the analogue of the hopfion 
would correspond to a Skyrmion, whose spectral and dynamical 
properties would be directly
accessible through our approach.
%
It would also be of interest to explore extensions in open systems, such as 
finite temperature BECs~\cite{prouk}, the (chiefly quasi-two-dimensional)
polariton superfluids~\cite{carusotto},  and damped-driven systems, more
generally. 
%

\begin{acknowledgments}
We thank A.J.~White for useful discussions.
W.W.~acknowledges support from NSF (grant No.~DMR-1208046).
P.G.K.~gratefully acknowledges the support of
NSF-DMS-1312856, of the ERC under FP7, Marie Curie Actions, People,
International Research Staff Exchange Scheme (IRSES-605096) and insightful
discussions with Profs.~I. Danaila and~B. Malomed.
R.C.G.~gratefully acknowledges the support of NSF-DMS-1309035.
The work of D.J.F.~was partially supported by the Special Account
for Research Grants of the University of Athens.
This work was performed under the auspices of the Los Alamos National
Laboratory, which is operated by LANS, LLC for the NNSA of the U.S.~DOE
under Contract No.~DE-AC52-06NA25396.
\end{acknowledgments}




\bigskip
\bigskip
\bigskip
\begin{center}
\bf
SUPPLEMENTAL MATERIAL:\\[1.0ex]
Robust Vortex Lines, Vortex Rings and Hopfions in 3D Bose-Einstein Condensates
\end{center}

\begin{figure}[ht]
\begin{center}
\includegraphics[width = 3.2in]{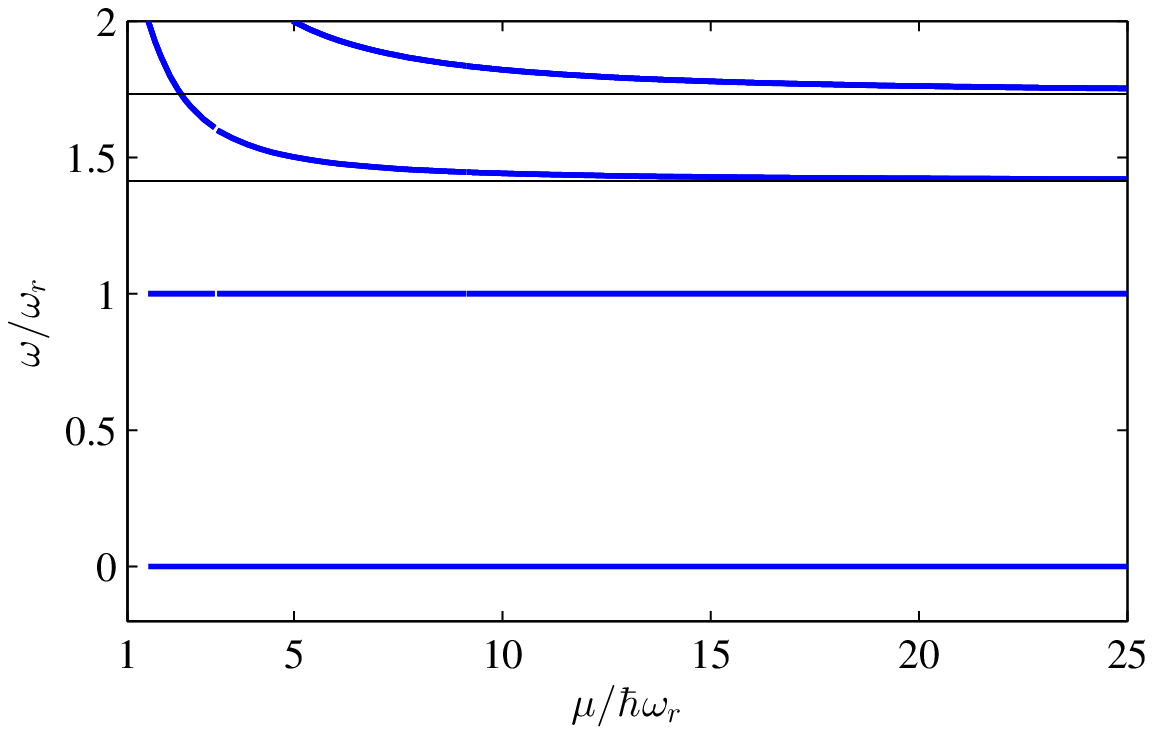}
\includegraphics[width = 3.2in]{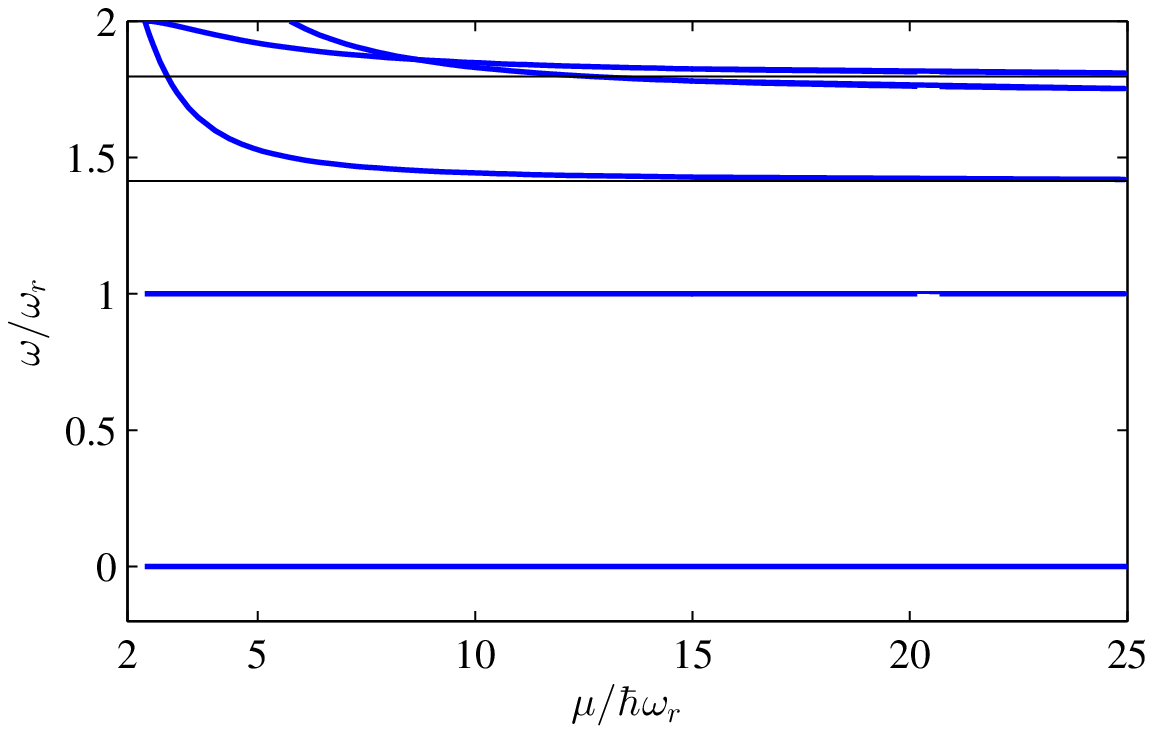}
\caption{
(Color online) BdG-frequency spectrum for the ground state 
(bearing no vorticity) when 
$\omega_z/\omega_r=1$ (top panel) and
$\omega_z/\omega_r=2.8$ (bottom panel). 
Stable (real) components are depicted by the thick (blue) lines.
There are no unstable (imaginary) eigenvalues for the ground state.
The thin horizontal lines pertain to the TF analytical
predictions of Ref.~\cite{Stringari1996}.
}
\label{Fig:GSfz1-GSfz2p8}
\end{center}
\end{figure}

\begin{figure}[ht]
\begin{center}
\includegraphics[width = 3.2in]{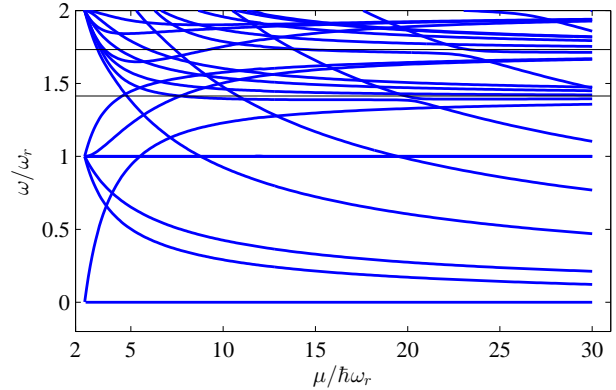}
\caption{
(Color online) BdG-frequency spectrum for the vortex line when $\omega_z/\omega_r=1$.
Stable (real) components are depicted by the thick (blue) lines.
There are no unstable (imaginary) eigenvalues for this case.
The thin horizontal lines pertain to the TF analytical
predictions of Ref.~\cite{Stringari1996} for the ground state modes.
}
\label{Fig:GSVL}
\end{center}
\end{figure}

\begin{figure}[ht]
\begin{center}
\includegraphics[width = 3.2in]{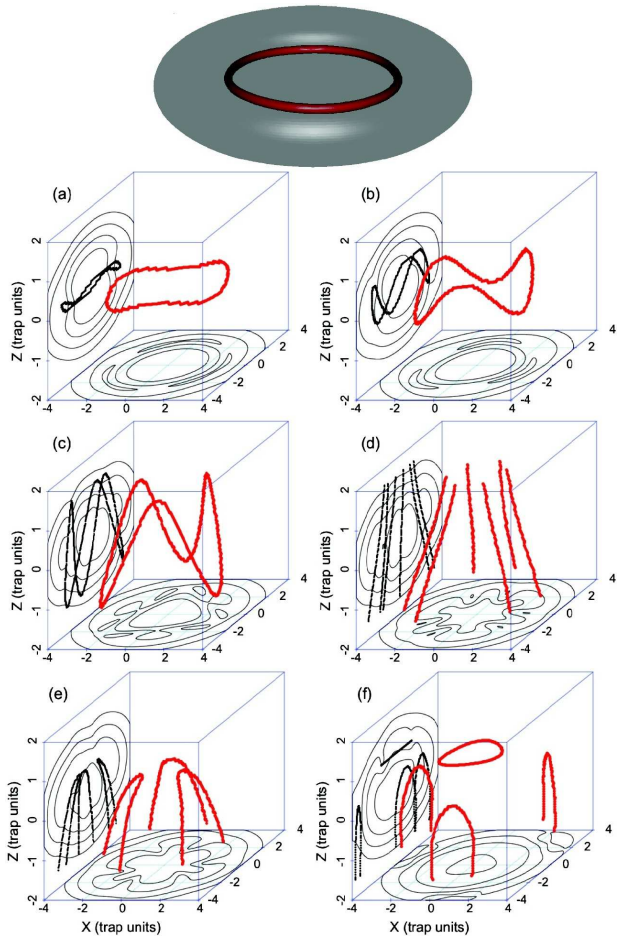}
\caption{
(Color online) Dynamics of the VR, demonstrating instability to an $n=3$ undulation, shown at times 
$t/\omega_r = $ 0 (a), 4.9 (b), 7.4 (c), 8.4 (d), 9.4 (e) and 11.6 (f). 
Parameters: $\omega_z/\omega_r=2.8$ and $\mu \approx 15.3$.
Red (gray) curves indicate vortex core positions; thick black lines show the vortex core projections onto the $(y,z)$ plane; thin-black contour lines represent the column density projections at 0.2, 0.4, 0.6 and 0.8 of the peak. Top: iso-density surface of the stationary state at 0.1 of the peak; the inner iso-density surface, due to the core, is colored red.
To see a movie of the dynamics see Ref.~\cite{movies} [Movie\#2].
}
\label{Fig:VRN3Dyn}
\end{center}
\end{figure}

Figure \ref{Fig:GSfz1-GSfz2p8} depicts the BdG spectrum for the
ground state of the system, i.e., in the absence of any
coherent structure, starting from the linear
limit and continuing all the way to the large $\mu$, so-called
Thomas-Fermi (TF) limit. The top panel depicts the spectrum
for an isotropic trap ($\omega_z/\omega_r=1$) while the bottom
panel corresponds to an oblate trap ($\omega_z/\omega_r=2.8$). 
The thick (blue) lines correspond to the numerical results, 
while the thin horizontal (black) lines correspond to the
first few collective mode excitations predicted by the 
theory~\cite{Stringari1996}.
It is clear 
that, as the chemical potential $\mu$
increases, corresponding to a higher number of atoms in the condensate,
the full numerical spectrum coincides with the collective
excitations prescribed by the theory.

Figure \ref{Fig:GSVL} depicts the BdG spectrum for the
vortex line (VL) in an isotopic trap ($\omega_z/\omega_r=1$).
Notice that modes corresponding to the ground state
(see Fig.~\ref{Fig:GSfz1-GSfz2p8}) are also contained in the 
spectrum of the VL and, as $\mu$ increases, one recovers again
the collective excitations prescribed by the theory~\cite{Stringari1996}.
In a similar manner as the spectrum for the VL contains
the spectrum of the background cloud,
the modes corresponding to:
(i) the collective excitations of the background cloud 
(see Fig.~\ref{Fig:GSfz1-GSfz2p8}),
(ii) the VL (see Fig.~\ref{Fig:GSVL}), and
(iii) the VR (see Fig.~1 in the main text)
are ``inherited'' by the hopfion  as it may be noted in
Fig.~2 of the main text.

Finally, in Fig.~\ref{Fig:VRN3Dyn} we depict the evolution of
the unstable dynamics for the $n=3$ undulation mode of a
VR in an oblate trap ($\omega_z/\omega_r=2.8$).
The dynamics is similar as the one observed for the unstable
evolution of the $n=2$ mode in Fig.~3 of the main text.
However, in this case, owing to the higher undulation 
number of the unstable mode, the dynamics is more complex and
involves the breakup of the VR into six filaments (non-straight VLs) 
traversing the BEC cloud. These filaments in turn interact and
reconnect forming other (smaller) VRs that in turn evolve
and interact with the other filaments.
To see movies corresponding to the destabilization of 
(a) the VR along the $n=2$ mode [Movie\#1],
(b) the VR along the $n=3$ mode [Movie\#2],
and
(c) the hopfion [Movie\#3], please see Ref.~\cite{movies}.

%
%
%
%
%

\end{document}